\documentclass[12pt]{article}
\usepackage{verbatim, amsmath, amssymb, booktabs, multirow, mathtools, caption, cprotect, float}
\usepackage{graphicx,psfrag,epsf}
\usepackage{enumerate}
\usepackage{natbib}
\usepackage{textcomp}
\usepackage[hyphens]{url} 
\usepackage{hyperref}

\newcommand{\blind}{0}

\usepackage{booktabs}
\usepackage{longtable}
\usepackage{array}
\usepackage{multirow}
\usepackage{wrapfig}
\usepackage{float}
\usepackage{colortbl}
\usepackage{pdflscape}
\usepackage{tabu}
\usepackage{threeparttable}
\usepackage{threeparttablex}
\usepackage[normalem]{ulem}
\usepackage{makecell}
\usepackage{xcolor}

\usepackage{graphicx, verbatim, amsmath, amssymb, booktabs, multirow, mathtools, caption, cprotect}

\begin{document}



\if0\blind
{
  \title{\bf Confounding and Regression Adjustment in Difference-in-Differences
Studies}

  \author{
        Bret Zeldow \\
     and \\     Laura A. Hatfield \\
    Department of Health Care Policy, Harvard Medical School\\
      }
  \maketitle
} \fi

\if1\blind
{
  \bigskip
  \bigskip
  \bigskip
  \begin{center}
    {\LARGE\bf Confounding and Regression Adjustment in Difference-in-Differences
Studies}
  \end{center}
  \medskip
} \fi

\bigskip
\begin{abstract}
Difference-in-differences (diff-in-diff) is a study design that compares
outcomes of two groups (treated and comparison) at two time points (pre-
and post-treatment) and is widely used in evaluating new policy
implementations. For instance, diff-in-diff has been used to estimate
the effect that increasing minimum wage has on employment rates and to
assess the Affordable Care Act's effect on health outcomes. Although
diff-in-diff appears simple, potential pitfalls lurk. In this paper, we
discuss one such complication: time-varying confounding. We provide
rigorous definitions for confounders in diff-in-diff studies and explore
regression strategies to adjust for confounding. In simulations, we show
how and when regression adjustment can ameliorate confounding for both
time-invariant and time-varying covariates. We compare our regression
approach to those models commonly fit in applied literature, which often
fail to address the time-varying nature of confounding in diff-in-diff.
\end{abstract}

\noindent%
\vfill

\newpage


\section{Introduction}

Difference-in-differences (diff-in-diff) studies contribute to policy
discourse by evaluating efficacy of newly enacted policies and programs.
For example, diff-in-diff has been used to estimate the effects of
raising minimum wage on employment rates \citep{card1994} as well as the
effects of new medical cannabis laws on opioid prescriptions
\citep{bradford_association_2018}. Diff-in-diff's most attractive
features are its simplicity and wide applicability; anyone with a
rudimentary understanding of experimental design and regression can
implement it. To carry out diff-in-diff, we just require observations
from a treated group and an untreated (comparison) group both before and
after the intervention is enacted.

Recent studies have leveraged diff-in-diff to estimate the effects of
expanded Medicaid eligibility through the Affordable Care Act (ACA) in
the United States. Following the ACA's passage and subsequent Supreme
Court ruling \citep{oyez}, each state chose whether to expand its
threshold for Medicaid eligibility. Some did and others did not,
creating groups of treated and comparison states and enabling natural
experiements using diff-in-diff \citep{kff}. For example, one of these
\citep{blavin_association_2016} showed that hospitals in states that
expanded Medicaid saw lower uncompensated care costs. Another showed
that people in Medicaid expansion states experienced improved access to
and affordability of health care \citep{kobayashi2019effects}. These
studies have informed ongoing policy debates about the future of the ACA
and state Medicaid waivers.

As in any causal inference prodcedure, diff-in-diff relies on strong and
unverifiable assumptions. The key assumption for diff-in-diff is that
the outcomes of the treated and comparison groups would have evolved
similarly \emph{in the absence of treatment}. Notably, diff-in-diff does
not require the treated and comparison groups to be balanced on
covariates, unlike in cross-sectional studies. Thus, a covariate that
differs by treatment group and is associated with the outcome is not
necessarily a confounder in diff-in-diff. Only covariates that differ by
treatment group and are associated with outcome \emph{trends} are
confounders in diff-in-diff as these are the ones that violate our
causal assumptions.

Despite the lurking pitfalls, many diff-in-diff studies appear to be run
on autopilot: plot the data, test for parallel outcome trends before the
intervention, and fit a regression that includes an interaction between
time with treatment, perhaps with some adjustment for covariates. Rarely
are the mechanisms of confounding discussed or the model specifications
interrogated.

In this paper, we discuss the unique features of diff-in-diff that run
afoul of our understanding of confounding and regression adjustment
imported from other settings. Confounders are fundamentally different in
diff-in-diff. We show how covariates, both time-invariant and
time-varying, affect the causal assumptions and inform analysis choices.
Using simulations, we demonstrate how to adjust for these confounders
and compare regression to matching techniques. We offer applied
researchers advice and strategies to estimate unbiased causal effects
using diff-in-diff by combining substance matter expertise with
thoughtful modeling.

\section{Parallel Trends}

In cross-sectional studies, the definition of a confounder comes from
the assumption that potential outcomes are independent of treatment.
Colloquially, we say that a confounder is a covariate associated with
both treatment and outcome, and we must condition on all confounders for
independence between treatment and outcomes to hold.
\citet{vanderweele_definition_2013} noted the lack of rigor in the
definition of a confounder and provided several formal definitions. In
this spirit, we examine what confounding means in diff-in-diff.

Diff-in-diff studies focus on the average effect of treatment on the
treated (ATT) at post-intervention point \(t^* \geq T_0\), where
\begin{equation}\label{eq:att}
ATT(t^*) = E\left\{ Y^1(t^*) - Y^0(t^*) \mid A = 1 \right\}\;,].
\end{equation} \noindent \(T_0\) is the time at which the policy is
implemented, \(A = 1\) represents the treated group, and \(Y(t)\) is a
continuous outcome recorded at time \(t\) with \(Y^a(t)\) denoting its
counterfactuals. Since Eq.~(\ref{eq:att}) contains counterfactuals we
never observe (that is, \(Y^0(t^*)\) for the treated group), we rely on
assumptions to identify this quantity using observables. To start, we
assume no anticipation effects of treatment so that the pre-treatment
outcomes are not affected by any treatment received in the future. From
this, it follows that the observed outcomes and the potential outcomes
are the same at pre-treatment times,
\(Y(t) = Y^0(t) = Y^1(t) \mbox{ for } t < T_0\). We also assume that the
post-treatment potential outcome corresponds to actual treatment
received, \(Y(t) = Y^0(t)(1-A) + Y^1(t)A\).

Identification relies on the parallel trends assumption, which we
formally define in the simplest possible setting of two time points, one
pre- and one post-treatment. Although some literature on diff-in-diff
separates the key assumption into two components, parallel trends and
common shocks \citep[Chapter 5.2]{angrist_mostly_2008}, we use the term
``parallel trends'' to refer to the combination of the two and write it
formally as \begin{equation}\label{eq:canon}
E\left\{ Y^0(1) - Y^0(0) \mid A = 0 \right\} = E\left\{ Y^0(1) - Y^0(0) \mid A = 1 \right\}\;.
\end{equation}

The assumption in Eq.~(\ref{eq:canon}) is based on \emph{changes} in
potential outcomes. That is, we assume the average change in the
untreated potential outcomes from pre- to post-treatment is the same for
the treated and comparison groups. Since the untreated potential outcome
in the post-treatment period \(\left(Y^0(1)\right)\) is unobservable for
the treated group \((A = 1)\), this assumption is untestable.

This definition of parallel trends with two time points is nearly
universal in the diff-in-diff literature
\citep{abadie_semiparametric_2005}. However, data in many applications
contain more than two time points, so we extend the assumption
accordingly. Let \(T\) be the total number of time points and
\(T_0 \leq T\) be the first post-treatment time point. In the strictest
version of parallel trends, every pair of time points satisfies
Eq.~(\ref{eq:canon}). That is, \begin{equation} \label{eq:long}
E\left\{ Y^0(t^*) - Y^0(t') \mid A = 0 \right\} = E\left\{ Y^0(t^*) - Y^0(t') \mid A = 1 \right\},
\end{equation} \noindent for \(t^* \neq t'\). While it is possible to
relax this assumption, this is the version researchers likely have in
mind when testing for parallel trends in the pre-intervention periods,
contending that evidence of parallel trends before treatment strengthens
the plausibility of parallel trends over the whole study period.

Given these assumptions and the parallel trends assumption in
Eq.~(\ref{eq:long}), we can re-write the ATT in a form that involves
only observable quantities \citep[Section 3.2.2]{lechner2011estimation},
as follows:

\begin{align} 
ATT(t^*) &= \left [ E\left\{ Y(t^*) \mid A = 1 \right\} -
  E\left\{ Y(t') \mid A = 1 \right\} \right ] - \nonumber\\ 
  &\;\; \left[ E\left\{ Y(t^*) \mid A = 0 \right\} -
  E\left\{ Y(t') \mid A = 0 \right\} \right ] \;, \nonumber
\end{align} \noindent with \(t' < T_0\). To estimate the ATT, we can now
select from a variety of estimators, ranging from a simple nonparametric
estimator using sample means to more sophiscated estimators such as
those using inverse probability weighting \citep{stuart_using_2014}.

\subsection{Regression Models for Difference-in-Differences}

We start by specifying a simple model for the untreated potential
outcomes conditional on a covariate. Following convention in
diff-in-diff literature \citep{oneill_estimating_2016}, we write a
linear model for the expected untreated potential outcomes of the
\(i^{th}\) unit \begin{equation} \label{eq:unt}
E\left[Y_i^0(t) | A = a_i, X = x_{it} \right] = \alpha_0 + \alpha_1 a_i + \zeta_t + \lambda_t x_{it}\;,
\end{equation} \noindent where \(\zeta_t\) are time fixed effects and
\(a_i\) is an indicator of the treated group (i.e, \(a_i = 1\) if the
\(i^{th}\) unit is in the treated group and \(a_i = 0\) otherwise). We
allow the covariate \(x_{it}\) to vary across units \(i\) and (possibly)
across time \(t\). Let \(\alpha_0\) be the intercept, \(\alpha_1\) the
constant difference between treated and comparison groups, and
\(\lambda_t\) the time-varying effect of the covariate on the outcome.
Denote the group-time mean of the covariate \(E[X_{it} | A = a]\) by
\(\tau_{a,t}\).

We pause here to note that there are a handful of other data-generating
models proposed in different settings. For example, \citet{bai2009panel}
proposes an interactive fixed effects model. The generalized synthetic
control method extends interactive fixed effects by adding heterogeneous
treatment effects \citep{xu2017generalized}. All this indicates that
there are many ways to set up this problem. We chose the above because
it is straightforward and familiar to most readers. However,
investigating the effect of confounding under different models may pose
unique challenges.

Assuming the data-generating model from Eq. (\ref{eq:unt}), we can
identify situations in which the covariate is a confounder for our
diff-in-diff estimator, meaning that the presence of the covariate
threatens the parallel trends assumption when not properly accounted
for. In the following sections, we show that, for a time-invariant
covariate, the parallel trends assumption will be violated (and \(X\)
will be a confounder) when two conditions hold: (1) the mean of \(X\)
varies by treatment group and (2) the relationship of \(X\) to the
outcome varies over time. For a time-varying covariate, \(X\) will be a
confounder if its distribution evolves differentially between the
treated and comparison groups (regardless of whether the effect on the
outcome is constant).

\subsection{Parallel Trends in the Presence of Covariates}

We demonstrate the conditions described above in the simple case of only
two time points, \(t \in \{0, 1\}\). We begin with expressions for the
mean change in untreated potential outcomes from pre- to post-treatment
in each group, by plugging Eq. (\ref{eq:unt}) into the parallel trends
assumption of Eq. (\ref{eq:canon}). In the treated group, the change
over time is \begin{align*}
E\left[Y^0(1) - Y^0(0) \mid A = 1\right] &= 
  (\alpha_0 + \alpha_1 + \zeta_1 + \lambda_1 \tau_{1,1}) - 
  (\alpha_0 + \alpha_1 + \zeta_0 + \lambda_0 \tau_{1,0}) \\
&= \zeta_1 - \zeta_0 + \lambda_1 \tau_{1,1} - \lambda_0 \tau_{1,0}\;,
\end{align*} \noindent and for the comparison group, it is
\begin{align*}
E\left[Y^0(1) - Y^0(0) \mid A = 0\right] &= 
  (\alpha_0 + \zeta_1 + \lambda_1 \tau_{0,1}) - 
  (\alpha_0 + \zeta_0 + \lambda_0 \tau_{0,0}) \\
&= \zeta_1 - \zeta_0 + \lambda_1 \tau_{0,1} - \lambda_0 \tau_{0,0}\;.
\end{align*} Subtracting the two, we get the differential change in
untreated potential outcomes between treated and comparison groups:
\begin{align}
\left(\zeta_1 - \zeta_0 + \lambda_1 \tau_{1,1} - \lambda_0 \tau_{1,0} \right) - 
\left(\zeta_1 - \zeta_0 + \lambda_1 \tau_{0,1} - \lambda_0 \tau_{0,0} \right) &= \nonumber \\ 
\lambda_1 \left(\tau_{1,1}-\tau_{0,1} \right) - \lambda_0 \left(\tau_{1,0} - \tau_{0,0}\right)\;. \label{eq:diff-change}
\end{align} The parallel trends assumption in Eq. (\ref{eq:canon})
constrains this difference to be 0. Given the data-generating model in
Eq. (\ref{eq:unt}), we can put conditions on the means and coefficients
of the covariates (\(\lambda\)'s and \(\tau\)'s) that will ensure the
parallel trends assumption holds. Then we define confounders as
variables that fail to satisfy those conditions.

First, consider a covariate that is constant over time (e.g., birth
year). Writing the mean of \(X\) in the treated group
\(\tau_{1,0} = \tau_{1,1} = \tau_1\) and in the comparison group as
\(\tau_{0,0} = \tau_{0,1} = \tau_0\), the differential change in Eq.
(\ref{eq:diff-change}) simplifies to
\begin{equation} \label{eq:simplified}
\lambda_1 (\tau_{1}-\tau_{0}) - \lambda_0 (\tau_{1}-\tau_{0}) = 
(\lambda_1 - \lambda_0) (\tau_{1}-\tau_{0})\;.
\end{equation} \noindent Whenever \(\tau_0 \neq \tau_1\),
Eq.~(\ref{eq:simplified}) will be zero if and only if
\(\lambda_0 = \lambda_1\). Conversely, if \(\lambda_0 \neq \lambda_1\),
Eq.~(\ref{eq:simplified}) will be zero if and only if
\(\tau_0 = \tau_1\). This implies that for a time-invariant covariate,
and absent the effects of other factors, parallel trends holds if
either: (1) the means of the covariate are the same across groups or (2)
the effect of the covariate on the outcome is the same across time
points.

Next, consider a covariate that varies over time (e.g., blood pressure
measured at each \(t\)). Eq.~(\ref{eq:diff-change}) will be zero ---
satisfying parallel trends --- if two conditions are met: the
relationship of the covariate to the outcome is constant
(\(\lambda_0 = \lambda_1\)) \emph{and} the difference in the mean of the
covariate between groups is equal
(\(\tau_{1,1} - \tau_{0,1} = \tau_{1,0} - \tau_{0,0}\)). From this, we
can see a time-varying covariate is a confounder if its relationship to
the outcome is time-varying \emph{or} the covariate evolves differently
in the treated and comparison groups.

Putting this all together, a confounder in diff-in-diff is a variable
with a time-varying effect on the outcome or a time-varying difference
between groups. Compare this to the colloquial defintion of a confounder
in cross-sectional settings: a variable associated with both treatment
and outcome. In diff-in-diff, a confounder always has some time-varying
effect. Either the relationship of the variable to the outcome changes
over time or the variable evolves differently between the groups over
time.

Next, we consider adjusting for these types of confounding variables in
the data-generating model of Eq.~(\ref{eq:unt}) using a linear
regression model in which we assume the confounder is measured. An
effective adjustment strategy must remove either covariate differences
between groups or account for their time-varying effects on the outcome.
In addition to regression adjustment, one might also consider matching
and inverse propensity score techniques
\citep{ryan_why_2015, stuart_using_2014}. We discuss matching briefly in
Section \ref{s:matching} and compare it to regression in Section
\ref{s:discussion}.

\section{Adjusting for Confounders}

To facilitate a regression approach for confounder adjustment, we first
connect the untreated potential outcomes in Eq.~(\ref{eq:unt}) to the
treated potential outcomes and then to the observed outcomes. First, we
assume a constant, additive effect of treatment, relating the treated
and untreated potential outcomes for post-treatment times \(t \geq T_0\)
as \[
Y_i^1(t) = Y_i^0(t) + \gamma\;.
\] Then we write the expected observed outcomes as
\begin{equation}\label{eq:obs-linear}
E\left[Y_i(t) \mid  A = a_i, X = x_{it}\right] = \alpha_0 + \zeta_t + \alpha_1 a_i + \lambda_t x_{it} + \gamma p_t a_i\;,
\end{equation} \noindent where \(p_t\) is an indicator of being in a
post-treatment time point. We use a linear regression model to estimate
the diff-in-diff parameter \(\gamma\).

\subsection{Adjusting for Time-Invariant Confounders}

Whenever \(X\) is a time-invariant baseline confounder and we use a
linear regression model to estimate the ATT, simply including a term for
the main effect of \(X\) (in addition to the usual \(a\) group effect, a
post-treatment indicator \(p_t\), and their interaction) will not
eliminate bias. Nevertheless, methods in the applied literature
consistently adjust for main effects of observed covariates
\citep{mcwilliams_changes_2014, rosenthal-pay-2016, desai-association-2016, roberts-changes-2018}. Likely these
choices are made out of habit rathen than with consideration to the
unique assumptions of diff-in-diff. While inclusion of covariates might
not harm estimates of the ATT, it might not be necessary.

We demonstrate that adjusting only for main effects is ineffective in
correctly non-parallel trends using a toy example with two time points.
Suppose we have a time-invariant covariate \(x_i\) with different means
in the two groups, \(E[X \mid A = 0] = \tau_0 = 0\) and
\(E[X \mid A = 1] = \tau_1 = 1\), and a time-varying effect with
\(\lambda_0 = 0\) and \(\lambda_1 = 1\). Because we are interested in
the covariate's effect on parallel trends --- which involve only the
untreated counterfactuals --- we include no treatment effect. This means
the observed outcomes and the untreated potential outcomes are equal, so
we can illustrate our points in observed data. Outcomes are generated
from Eq.~(\ref{eq:unt}) with \(\alpha_0 = 1\), \(\alpha_1 = -1\),
\(\zeta_0 = 1\), and \(\zeta_1 = 2\). The covariate is a confounder
because its relationship to the outcome varies over time
(\(\lambda_0 \neq \lambda_1\)) and its means in the treated and
comparison groups differ (\(\tau_0 \neq \tau_1\)).

\begin{figure}
  \includegraphics[width=\linewidth]{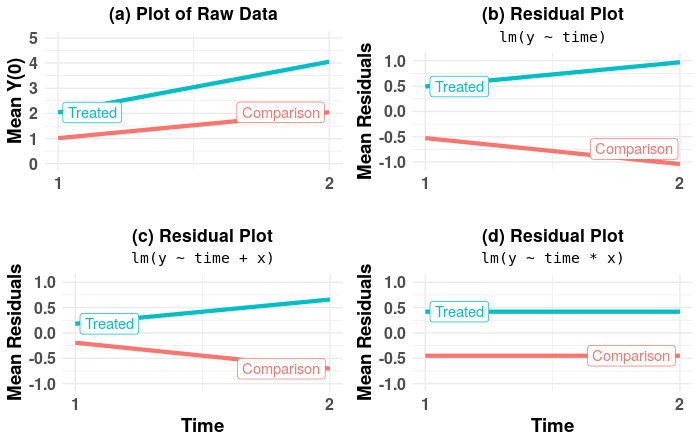}
  \cprotect\caption{Adjusting for the main effect of a covariate does not correct for diverging trends, but adjusting for the interaction with time does in a simulated example where untreated potential outcomes depend on a time-invariant covariate with a time-varying effect. Panel (a) shows untreated potential outcomes. Panels (b)-(d) show residuals from linear models. The function \verb|lm| fits a linear model with outcome \verb|y|. In panel (b), the only predictor is time. In panel (c), the predictors are the main effects of time and the covariate. In panel (d), the predictors are time, the covariate, and an interaction between the two. } \label{fig:adjust}
\end{figure}

In Panel (a) of Figure \ref{fig:adjust}, we plot the mean outcomes by
group and time. The non-parallel outcome evolution in the two groups is
apparent. Without accounting for the confounding, we would incorrectly
attribute differential outcome changes to the treatment. Panel (b) shows
residuals from a simple linear regression with only a time effect. This
model does not include the covariate \(X\), so we would not expect the
model to correct for deviations from parallel trends. We see that the
residuals, like the outcomes, are not parallel. In Panel (c), we add a
main effect for the covariate \(X\) to the model. However, the residuals
for the two groups still diverge. In Panel (d), we add an interaction
between \(X\) and time. Only in this model do we properly account for
the time-varying nature of the confounder and obtain an unbiased result
(recall the true treatment effect is zero here).

This illustrates just one data-generating scenario and a few simple
models. In the simultations of Section \ref{s:simulations}, we provide a
more comprehensive look at how covariate adjustment through regression
and matching can address confounding in diff-in-diff.

\subsection{Adjusting for Time-Varying Confounders}

Like time-invariant confounders, time-varying confounders invalidate
parallel trends and introduce bias into our estimate of the ATT. If we
were to adjust for time-varying confounding either by including the main
effect or its interaction with time in a regression, we risk
conditioning on post-treatment covariates that may be affected by
treatment. As \citet{rosenbaum1984consequences} notes for observational
data, at best adjusting for post-treatment covariates provides no
benefit; at worst, it may introduce additional bias. This is because the
time-varying covariate can act as both a confounder and as a mediator.
As such, when trying to recover the ATT via regression, the usual
interaction parameter may not be an unbiased estimate of the ATT.

To see why this is the case, imagine three different scenarios: (a) the
time-varying covariate changes in a way completely unrelated to
treatment, (b) the time-varying covariate changes in a way wholly
determined by treatment, and (c) the time-varying covariate changes in a
way determined by a combination of treatment and other factors. Whenever
(b) or (c) is true and the time-varying covariate is a cause of the
outcome, the ATT is a combination of the direct effect of treatment and
the indirect effect of treatment via the covariate. As a result the
regression parameter on the interaction between treatment and the
post-treatment indicator may not equal the ATT, even adjusted for the
time-varying covariate. However, if we fail to account for the
covariate, we face parallel trends violations. For a more detailed
explanation, please see Appendix Section A.

In the causal inference literature, g-methods were specifically designed
to deal with time-varying confounding \citep{hernan2019causal}. A
handful of papers incorporate these techniques into the diff-in-diff
framework such as inverse probability weighting
\citep{stuart_using_2014, han_evaluating_2017}. However, only
one employs inverse probability weighting to account for changes in
covariate distributions across time \citep{stuart_using_2014}. In that
paper, the authors consider a two time point/two group setting and
define a new variable with four levels (treatment group in the
pre-treatment period, treatment group in the post-treatment period,
etc.). However, this methodology was only demonstrated on data with two
time points and it should be noted that the target estimand changes from
the classic ATT to an average treatment effect defined in the treatment
group at the first time point. Nevertheless, it remains one of the only
diff-in-diff papers to directly address the issue of time-varying
confounders. In this paper, we use simulations to demonstrate that the
estimate of the ATT is biased when time-varying covariates are affected
by treatment, whether we adjust for the time-varying covariate or not
(see Scenario 6 of Section \ref{ss:time-varying}).

\subsection{What about Matching?}\label{s:matching}

Matching aims to reduce confounding bias by selecting units from the
treated and comparison groups that have similar observable
characteristics. This eliminates imbalances between the groups, which is
a key ingredient in confounding. When matching, we can match
observations on pre-treatment outcomes, pre-treatment covariates, or
some combination.

Matching on pre-treatment outcomes allows us to use an alternative
assumption to identify the target parameter. This assumption ---
independence between potential outcomes and treatment assignment
conditional on past outcomes --- is the basis of lagged dependent
variables regression and synthetic control methods
\citep{lechner2011estimation, oneill_estimating_2016, ding2019bracketing}.
However, matching on pre-treatment outcomes in diff-in-diff can yield
unwanted results. In some settings, it reduces bias
\citep{stuart_using_2014, ryan_why_2015}, while in others, matching
induces regression to the mean and \emph{creates} bias
\citep{oneill_estimating_2016, daw_matching_2018}.

Matching only on time-invariant pre-treatment covariates is attractive
because it removes differences in the covariate distribution between the
groups. With time-varying covariates, the picture is more complicated.
Matching on time-varying pre-treatment covariates is subject to the same
threat of bias due to regression to the mean as matching on
pre-treatment outcomes. Moreover, if confounding arises because of
differential evolution of the covariate in the two groups, matching only
on pre-treatment values will be insufficient to address the confounding.
Thus, we may wish to match on both pre- and post-treatment values of a
time-varying covariate. In this case, we must also be wary of the
dangers of matching on post-treatment variables that may be affected by
treatment \citep{rosenbaum1984consequences}. Clearly, choosing the right
matching variables is the key to effective matching. A good overview on
the current state of matching for diff-in-diff is provided by
\citet{lindner_difference_differences_2018}.

Returning to the demonstration of parallel trends in Figure
\ref{fig:adjust}, matching on the pre-treatment covariate also serves to
fix diverging trends. Recall that the data-generating model was a
time-invariant covariate with a time-varying effect on the outcome.
Eliminating the difference between the covariate means in the treated
and comparison group via matching is sufficient to address confounding.
If the confounding had arisen due to a time-varying covariate, the
strategy may not suffice.

Both matching and regression adjustment have potential pitfalls. In
addition to the possible regression to the mean problem mentioned above,
we can mistakenly match on noise or on a set of covariates that is
insufficient to alleviate bias in our causal effect. Furthermore,
matching choices are largely \emph{ad hoc} and can depend on the data
structure itself. For example, it's much more straightforward to match
in panel data than in repeated cross-sections. Regression adjustment is
not without its limitations as well. We can overfit our model, for one.
We can also choose the wrong covariates to include or mispecify the
functional form of the model. Deciding whether to address diverging
trends through matching or regression or both must be done carefully.
For example, say we are missing a key covariate that we suspect drives
divergent trends, we cannot address the bias through regression
adjustment and could instead consider matching on pre-treatment outcomes
as a proxy for the missing covariate. On the other hand, if we have
repeated cross-sectional data and it's not clear how to match
effectively, we can choose regression adjustment.

\section{Simulations}\label{s:simulations}

We use simulation to compare regression adjustment and matching
strategies in diff-in-diff. In each simulation scenario, we generate 400
datasets of \(n = 800\) units observed at \(T = 10\) time points. The
first 5 time points are pre-treatment times, and the last 5 are
post-treatment. Each unit is assigned to the treatment group with
probability 0.5. To each simulated data set, we apply regression and
matching techniques that reflect current practice in the applied
literature and compare the bias of the resulting treatment effects.

We simulate data and analyze it using the R environment (R version 3.6.1
\citep{r-lang}). We fit regression models using the \verb|lm| function
and estimate post-hoc cluster-robust standard errors using the
\verb|cluster.vcov| function in the \verb|multiwayvcov| package
\citep{graham2016multiwayvcov}. To match, we use the \verb|MatchIt|
package \citep{matchit}. We present averages, across simulated data
sets, of the percent bias and standard error of the estimated treatment
effect.

Below, we describe the specifics of our data-generating and analysis
models, first for scenarios with time-invariant covariates and then for
scenarios with time-varying covariates.

\subsection{Time-Invariant Covariate}

\subsubsection{Data-generating models}

\begin{table}
\begin{tabular}{p{0.3\textwidth}l}  
\toprule
Scenario    & Data-Generating Model \\
\midrule
\multirow{2}{3.5cm}{1: Time-invariant covariate effect} &  $X_i \stackrel{ind}{\sim} N\left(m(a_i), v(a_i)\right)$ \\
    & $Y_i(t) \stackrel{ind}{\sim} N\left(1 + a_i + a_i p_t + u_i + x_i + f(t), 1\right)$ \\
\midrule
\multirow{2}{3.5cm}{2: Time-varying covariate effect}   &  $X_i \stackrel{ind}{\sim} N\left(m(a_i), v(a_i)\right)$ \\ 
    & $Y_i(t) \stackrel{ind}{\sim} N\left(1 + a_i + a_i p_t + u_i + x_i + f(t) + g(x_i,t), 1\right)$ \\
\midrule
\multirow{3}{3.5cm}{3: Treatment-independent covariate} & $X_i \stackrel{iid}{\sim} N\left(1,1\right)$ \\ 
    & $Y_i(t) \stackrel{ind}{\sim} N\left(1 + a_i + a_i p_t + u_i + x_i + f(t) + g(x_i,t), 1\right)$ \\
    & \\
\bottomrule
\end{tabular}
\caption{Data-generating models with a time-invariant covariate, $x_i$. $Y_i(t)$: outcome for $i^{th}$ subject at time $t$. $a_i$: group indicator. $u_i$: random intercept. $p_t$: indicator of post-treatment time point. The treatment assignment for all scenarios is $a_i \stackrel{iid}{\sim} \text{Bernoulli}(0.5)$ and the unit-level intercepts are all $u_i \stackrel{iid}{\sim} N(0,.25^2)$. $m(a_i) = 1.5 - 0.5 a_i$, $f(t) = (t - 2.5)^2/10$, $g(x_i,t) = (x_i \cdot t)/10$, $v(a_i) = (1.5 - 0.5 a_i)^2$.} \label{tab:time-invariant-generating}
\end{table}

Our first set of simulations involves a time-invariant covariate. In
Scenario 1, the distribution of \(X\) is different in the treated and
control groups, but \(X\) has a time-invariant effect on the outcome
\(Y\). Scenario 2 is the same as Scenario 1 but we allow the effect of
\(X\) on \(Y\) to vary over time. In Scenario 3, the effect of \(X\) on
the mean of \(Y\) is again time-varying, but the distribution of \(X\)
is the same in the treated and control groups. Table
\ref{tab:time-invariant-generating} summarizes the data-generating
processes for these three simulations

We expect that in Scenarios 1 and 3, anlyses that do not adjust for
\(X\) will be unbiased, because \(X\) does not satisfy the definition of
a confounder. In Scenario 1, this is because \(X\) does not have a
time-varying effect on \(Y\); in Scenario 3, this is because the
distribution of \(X\) is the same in both groups. In Scenario 2, we
expect that only analyses that adjust appropriately for the time-varying
effect of \(X\) on \(Y\) will yield unbiased results. For all three
scenarios, the ATT equals the regression parameter which was set to 1.
We measure bias with respect to this true ATT.

\subsubsection{Analysis approaches}\label{sss:time-invariant-analysis}

We use both matched and unmatched regression to analyze the simulated
data. All regression models include time fixed effects and indicators
for treatment, the post-period, and their interaction. The simple model
includes only those elements, ignoring the covariate entirely. The
covariate adjusted (CA) model adjusts for the covariate using a constant
effect on the outcome over time. The time-varying adjusted (TVA) model
allows the coefficient on the covariate to vary over time.

Our matching strategies include matching on both outcomes and
covariates. We use nearest-neighbor matching on 1) the vector of
pre-treatment outcomes, 2) the vector of pre-treatment outcome first
differences, or 3) pre-treatment covariates. To each matched dataset, we
fit a simple model without covariate adjustment. Table
\ref{tab:sim-models} describes the adjustment methods and gives pseudo
code for each.

\begin{table}
\begin{tabular}{ll}
\toprule
Model & Pseudo \verb|R| code \\
\midrule
Simple &
\begin{minipage}{2.9in}
\begin{verbatim}
lm(y ~ a*p + t)
\end{verbatim}
\end{minipage}\\
Covariate-Adjusted (CA) &
\begin{minipage}{2.9in}
\begin{verbatim}
lm(y ~ a*p + t + x)
\end{verbatim}
\end{minipage}\\
Time-Varying Adjusted (TVA) &
\begin{minipage}{2.9in}
\begin{verbatim}
lm(y ~ a*p + t*x)
\end{verbatim}
\end{minipage}\\
Match on pre-treatment outcomes &
\begin{minipage}{2.9in}
\begin{verbatim}
lm(y ~ a*p + t,data=out.match)
\end{verbatim}
\end{minipage} \\
Match on pre-treatment first differences &
\begin{minipage}{2.9in}
\begin{verbatim}
lm(y ~ a*p + t,data=out.lag.match)
\end{verbatim}
\end{minipage} \\
Match on pre-treatment covariates & 
\begin{minipage}{2.9in}
\begin{verbatim}
lm(y ~ a*p + t,data=cov.match)
\end{verbatim}
\end{minipage} \\
\bottomrule
\end{tabular}
\cprotect\caption{Analysis methods applied to each simulation scenario. The function \verb|lm| fits a linear model for outcome \verb|y|, treatment group indicator \verb|a|, post-treatment period indicator \verb|p|, (factor-coded) time \verb|t|, and covariate \verb|x|. The notation \verb|p*q| yields main effects for \verb|p| and \verb|q| plus their interaction.}\label{tab:sim-models}
\end{table}

\subsection{Time-Varying Covariate}\label{ss:time-varying}

\subsubsection{Data-generating models}

The second set of simulations involves a time-varying covariate, with
means that may evolve differently in the treated and comparison groups.
The basic setup of these simulations (i.e., the number of units, time
points, and treatment assignment) is the same as in Scenarios 1 through
3 above. We include three types of covariate evolution. In Scenario 4,
the covariate evolves the same for both the treated group and the
comparison group; in Scenario 5, the covariate evolves differently
starting from baseline (related to treatment \emph{group}, not treatment
itself); and in Scenario 6, the covariate evolves the same in the two
groups before treatment but differently after treatment.

For all these scenarios, we have two outcome processes: (a) the
covariate has a time-invariant effect of the outcome and (b) the
covariate has a time-varying effect on the outcome. Each scenario embeds
two sub-scenarios, for a total of six data-generating processes. The
data-generating distributions are summarized in Table
\ref{tab:time-varying-generating}. For Scenarios 4 and 5, the ATT equals
the regression parameter (set to 1) as it did in Scenarios 1 through 3.
However, Scenario 6 has a covariate that is changed by treatment, acting
in part as a mediator. Thus, for Scenario 6, the ATTs are 0.85 and 0.87
for outcome processes (a) and (b), respectively. Work showing these
calculations is provided in Appendix Section B. For all scenarios, we
measure bias relative to the true ATT.

\begin{table}
\begin{tabular}{p{0.3\textwidth}ll}  
\toprule
Scenario    & Data-Generating Model \\
\midrule
\multirow{2}{3.5cm}{4: Parallel evolution} &  $X_{ti} = x_{(t-1)i} + m_1(t)\cdot z$ \\
    & $Y_i(t) \stackrel{ind}{\sim} N\left(1 + a_i + a_i p_t + u_i + x_{ti} + f(t) + g(x_i,t), 1\right)$ \\
\midrule
\multirow{2}{3.5cm}{5: Evolution differs by group}  & $X_{ti} = x_{(t-1)i} + m_2(a_i,t)\cdot z$ \\ 
    & $Y_i(t) \stackrel{ind}{\sim} N\left(1 + a_i + a_i p_t + u_i + x_{ti} + f(t) + g(x_i,t), 1\right)$ \\
\midrule
\multirow{2}{3.5cm}{6: Evolution diverges in post} & $X_{ti} = x_{(t-1)i} + m_1(t)\cdot z - m_3(a_i,t)$ \\ 
    & $Y_i(t) \stackrel{ind}{\sim} N\left(1 + a_i + a_i p_t + u_i + x_{ti} + f(t) + g(x_i,t), 1\right)$ \\
\bottomrule
\end{tabular}
\caption{Data-generating models with time-varying covariates, $x_{ti}$. $Y_i(t)$: outcome for $i^{th}$ subject at time $t$. $a_i$: group indicator. $u_i$: random intercept. $p_t$: indicator of post-treatment time point. The treatment assignment for all scenarios is 
$a_i \stackrel{iid}{\sim} \text{Bernoulli}(0.5)$, the unit-level intercepts are 
$u_i \stackrel{iid}{\sim} N(0,.25^2)$, the covariate value at the first time point is
$X_{1i} \stackrel{ind}{\sim} N\left(1.5-0.5 a_i,(1.5-0.5a_i)^2\right)$, and 
$z \stackrel{iid}{\sim} N(1,0.1^2)$. 
The functions that govern the evolution of the covariate are
$m_1(t) = (t-1)/10$, 
$m_2(a_i,t) = (I_{a_i=1}-I_{a_i=0})(t-1)/10$, and 
$m_3(a_i,t) = a_i p_t  t/20$. 
Outcome process (a) uses $f(t) = (t - 2.5)^2/10$ and $g(x_i,t)= 0$. 
Outcome process (b) uses $f(t) = (t - 2.5)^2/10$ and $g(x_i,t) = (x_i \cdot t)/10$.
} \label{tab:time-varying-generating}
\end{table}

\subsubsection{Analysis approaches}

The analysis methods are the same as in Section
\ref{sss:time-invariant-analysis} and Table \ref{tab:sim-models}, except
that for these scenarios with a time-varying covariate, we match on the
\emph{vector} of pre-treatment covariate values.

\subsection{Simulation Results}

\subsubsection{Time-Invariant Covariate}

\begin{figure}
\centering
\includegraphics{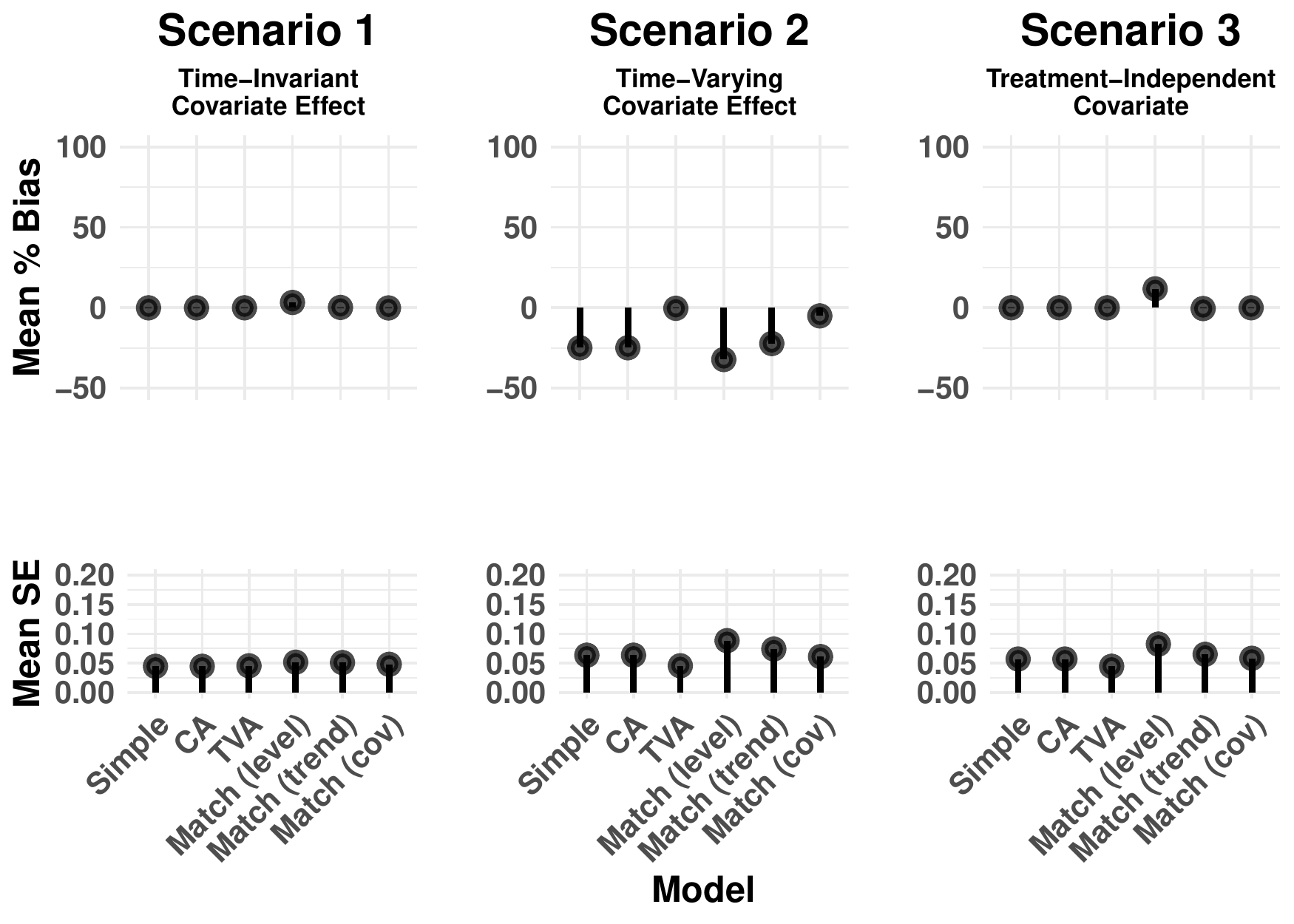}
\caption{\label{fig:time-invariant-results}Simulation Results for
Time-Invariant Covariate. CA = Covariate adjusted; TVA = Time-varying adjusted.}
\end{figure}

Figure \ref{fig:time-invariant-results} shows the results of fitting the
models in Table \ref{tab:sim-models} to the data generated from the
time-invariant covariate data-generating models in Table
\ref{tab:time-invariant-generating}. In Scenario 1, while \(X\) is
associated with treatment, it is not a confounder because the effect
does not vary over time. Thus, the unadjusted analysis (simple model) is
unbiased and adjusting for \(X\) in the CA and TVA models does not
affect either bias or standard errors. The results from our matched
regressions are similar to those from the unmatched regressions.

In Scenario 2, the time-varying effect of \(X\) on \(Y\) makes \(X\) a
confounder and thus requires covariate adjustment with a time-varying
aspect. Adjusting for the main effect of \(X\) (CA model) does not
alleviate bias or reduce the estimate's standard error. Fortunately, we
can address the bias by adjusting for the interaction of \(X\) with time
(TVA model). Of the matching strategies, only matching on the covariate
effectively eliminates bias.

In Scenario 3, the simple model is already unbiased because \(X\) is not
a confounder. In fact, all estimation strategies yield unbiased
estimates except matching on pre-treatment outcomes, which is biased by
about 10 percent due to regression to the mean. We see about 20\% lower
mean standard error when we adjust for the covariate in the TVA model
compared to the simple model.

\subsubsection{Time-Varying Covariate}

\begin{figure}
\centering
\includegraphics{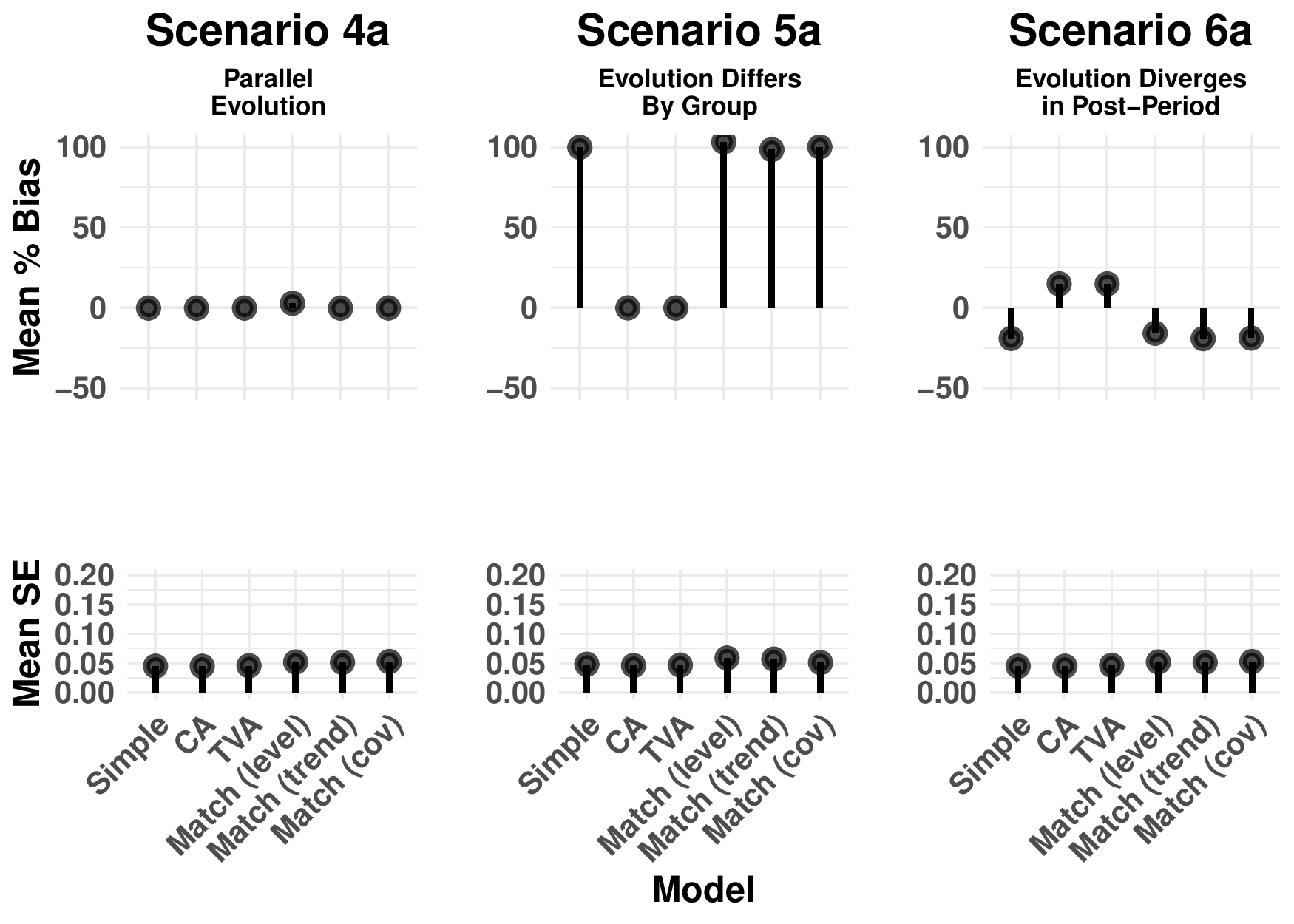}
\caption{\label{fig:time-varying-results}Simulation Results for
Time-Varying Covariate. CA = Covariate adjusted; TVA = Time-varying adjusted.}
\end{figure}

\begin{figure}
\centering
\includegraphics{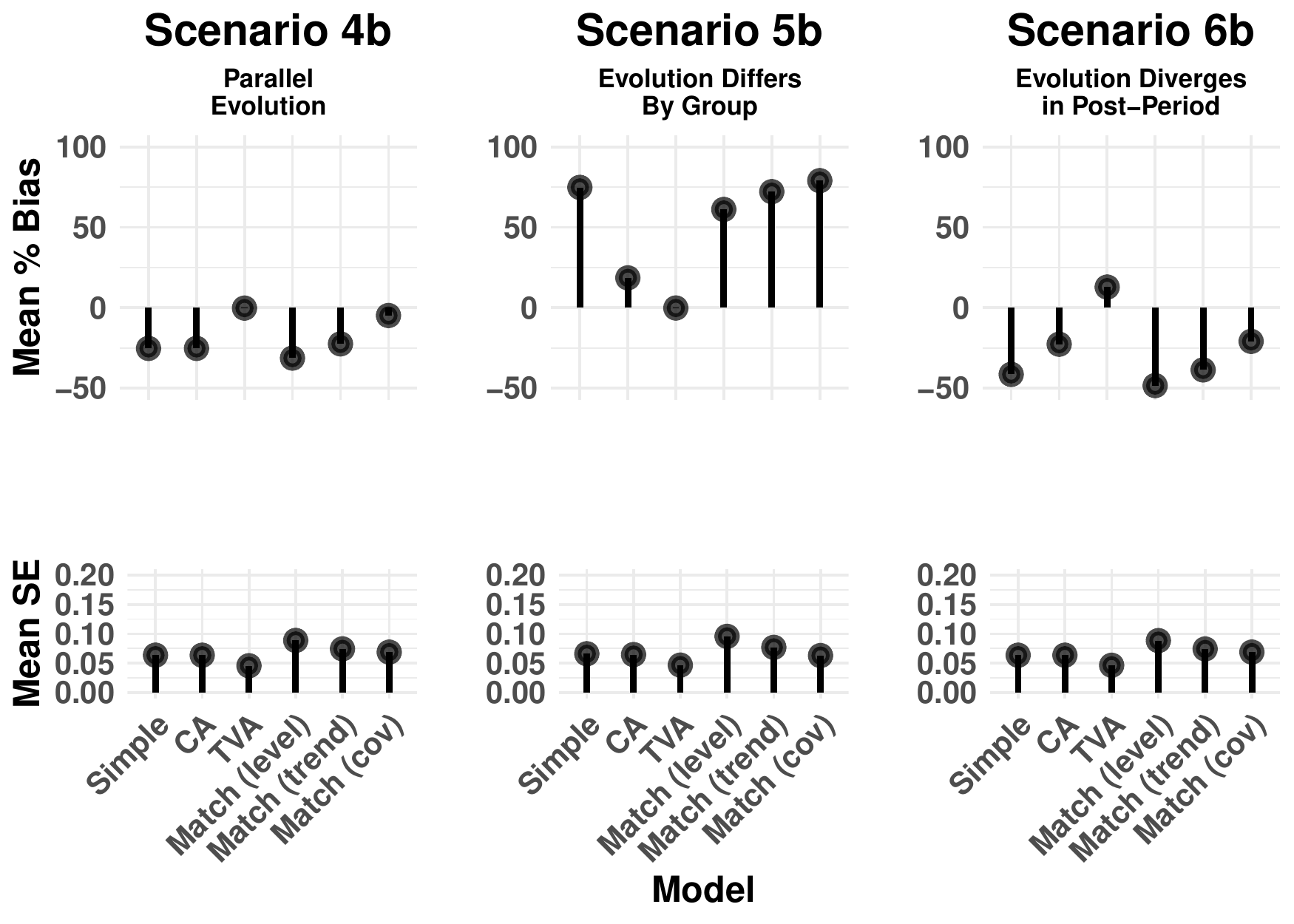}
\caption{\label{fig:time-varying2-results}Simulation Results for
Time-Varying Covariate. CA = Covariate adjusted; TVA = Time-varying adjusted.}
\end{figure}

Figures \ref{fig:time-varying-results} and
\ref{fig:time-varying2-results} show the results of fitting the models
in Table \ref{tab:sim-models} to the data generated using time-varying
covariate processes (Table \ref{tab:time-varying-generating}). In
Scenario 4, there is no confounding when the effect of \(X\) on \(Y\) is
constant over time, and the mean of \(X\) evolves the same for each
group. As a result, each modeling strategy is unbiased. However, when
\(X\) has a time-varying effect on \(Y\), \(X\) is a confounder and only
time-varying adjustment (TVA) eliminates bias. Matching on the vector of
pre-treatment values of \(X\) nearly eliminates the bias.

In Scenario 5, the time-varying covariate evolves differently by group,
beginning at baseline. When the effect of \(X\) on the outcome is
constant, we can simply adjust for time-varying \(X\) (CA model) to
eliminate confounding bias. When the effect of \(X\) on \(Y\) varies
over time, we must adjust for the interaction of \(X\) and time (TVA
model). All of the matching strategies have significant bias.

In Scenario 6, the time-varying covariate evolves differently by group,
but only \emph{after} the treatment is introduced at \(t = 6\). Recall
that in this scenario, the ATT does not simply equal the regression
coefficient on an interaction term. As a result in Scenario 6, we have
significant bias in our estimates and never succeed in recovering the
true ATT.

\section{Discussion} \label{s:discussion}

Diff-in-diff applications and methods have expanded dramatically over
the past few decades. We contribute to this growing literature by
examining how observable covariates may violate causal assumptions and
comparing regression strategies to adjust for violations. It is tempting
to toss all observed covariates into a regression model, but the form of
the model specification should be tailored to address time-varying
confounding.

Our methods and conclusions have several limitations. First, adjusting
for confounders spends degrees of freedom, which may be untenable for
sparse data. Second, regression adjustment depends on knowing and
measuring the confounders as well as the functional form of their
effects on the outcome (or having sufficient data to model it flexibly).
Third, our conclusions only apply to linear models; nonlinear models are
more complicated \citep{karaca-mandic_interaction_2012}.

Done properly, regression adjustment can address bias caused by
diverging trends. Further, even in the absence of confounding, adjusting
for covariates can improve efficiency of the effect estimate (see
Scenario 3 of Figure \ref{fig:time-invariant-results}). And a correctly
specified regression approach avoids conditioning on pre-treatment
outcomes and so is not susceptible to regression to the mean in the same
way that some matching methods are \citep{daw_matching_2018}. Lastly,
our regression adjustment strategy is agnostic to the structure of the
data, whether panel data versus repeated cross-sections. Our simulations
assumed panel data but our results will hold for repeated
cross-sections. Matching on repeated cross-sections is trickier, since
some covariates will necessarily be measured on different subjects at
different time points, but it is possible \citep{keele2019patterns}.

For researchers using diff-in-diff in applied work, we recommend several
steps for addressing confounding. First, researchers should clearly
specify their model and explain how the inclusion of covariates and
their functional forms support the researcher's assumptions and model.
This begins with writing out the full model specification and by
providing analysis code in supplementary materials. Each covariate and
coefficient should correspond to a threat to the validity of parallel
trends and provide a valid remedy. We also recommend researchers
comprehensively list covariates --- both observed and unobserved ---
that might cause violations of parallel trends. The list should contain
information on whether the variable is observed, whether the
distribution of the covariate is expected to differ in the treatment and
comparison groups, whether the covariate is time-varying, and whether it
has an effect on the outcome. Depending on the application, we can use
such a list to inform analysis choices. For example, if many unobserved
covariates are a concern, the analyst may choose a different estimator
(instead of one that relies on diff-in-diff and the parallel trends
assumption). On the other hand, a single time-invariant covariate
suggests a straightforward regression approach. Approaching both
measured and unmeasured covariates illuminates the crucial causal
assumptions underlying diff-in-diff more so than any test of parallel
pre-treatment outcomes \citep{bilinski_seeking_2018}. Other authors have
given similar advice, stressing attention to the reasons for baseline
differences between the treated and comparison groups and how these
differences might affect parallel trends \citep{kahn-lang_promise_2018}.

Being thorough in our diff-in-diff studies will strengthen conclusions
and help alleviate concerns on the credibility of parallel trends. We
expect diff-in-diff to continue its critical role in informing policy
decisions into the foreseeable future. Going forward, it is crucial that diff-in-diff methodology is developed with input from
statisticians, epidemiologists, economists, political scientists, and
policy analysts alike.

\section*{Acknowledgements}

The authors thank Alyssa Bilinski for helpful comments on the draft. This work was supported by funding from the Laura and John Arnold Foundation. The content is solely the responsibility of the authors and does not necessarily represent the views of the Laura and John Arnold Foundation.

\bibliographystyle{agsm}
\bibliography{HPDSLab}

\setcounter{table}{0}
\renewcommand{\thetable}{A\arabic{table}}

\section*{Appendix A - Adjusting for Time-Varying Covariates}

In this section of the appendix, we discuss of the problems of adjusting for time-varying confounders as described in Section 3.2 in the main paper. The thesis of our argument is that a time-varying covariate that is *affected* by treatment and also affects the outcome makes recovering the causal effect difficult. On one hand, failing to adjust for the time-varying covariate will result in failures of parallel trends. On the other hand, adjusting for the time-varying covariate, since it is on the pathway between treatment and the outcome, will negate some of the effect of treatment on the outcome, resulting in biased estimates.

We begin with notation that should be familiar to those who read our paper. $Y(t)$ is the continuous outcome measured at time $t$. For simplicity, we assume that $t \in \{0, 1\}$ where $t = 0$ is the pre-treatment period and $t = 1$ the post-treatment period. Treatment is binary and represented by $A$. Finally, we have a time-varying covariate $X_{it}$ where $i$ in an index for a unit (e.g., a state or an individual). Let $\tau_{at} = E[X_{it} \mid A = a]$ be the covariate group-time mean. We also introduce counterfactual notation for the covariate so that $X^a_{it}$ is the (possibly counterfactual) value of $X$ for individual $i$ and time $t$ under treatment $A = a$. Since we assume that treatment directly affects $X$, we may have that $X^0_{i1} \neq X^1_{i1}$.

Let's extend the notation for the covariate means to counterfactual world so that $E(X^0_{it} \mid A = a) = \tau^0_{at}$ and $E(X^1_{it} \mid A = a) = \tau^1_{at}$. We assume that treatment (which occurs between times 0 and 1) does not affect past versions of $X$ so that $\tau^0_{a0} = \tau^1_{a0} = \tau_{a0}$. We also assume that the covariate evolves differently in the two groups even absent treatment, leading to the failure of parallel trends. That is, $\tau_{01} - \tau_{00} \neq \tau_{11} - \tau_{10}$.

Suppose we have the same model for untreated outcomes as the main text:

\[
E\left[Y_i^0(t) | A = a_i, X^0 = x^0_{it} \right] = \alpha_0 + \alpha_1 a_i + \zeta_t + \lambda_t x^0_{it}.
\]
For simplicity, let $\lambda_t = \lambda$. We can connect the untreated outcomes to the treated outcomes with a fixed treatment effect, $\gamma$: $Y_i^1(t) = Y_i^0(t) + \gamma\;.$

Recall that the average treatment effect on the treated (ATT) is

\[
ATT = E\left\{ Y^1(1) - Y^0(1) \mid A = 1 \right\}\;.
\]

Now, we have:

\[
E\left\{Y^1(1) \mid A = 1\right\} = \alpha_0 + \alpha_1 + \zeta_1 + \lambda_1 \tau^1_{11} + \gamma
\]

and

\[
E\left\{Y^0(1) \mid A = 1\right\} = \alpha_0 + \alpha_1 + \zeta_1 + \lambda_1 \tau^0_{11}.
\]

Plugging into the ATT:

\[
ATT = \lambda_1 \tau^1_{11} + \gamma - \lambda_1 \tau^0_{11} = \gamma + \lambda_1(\tau^1_{11} - \tau^0_{11}).
\]

The ATT is what we want to calculate, but what is our estimate for an unadjusted model versus one from a regression model that correctly adjusts for $X$. \\

\vspace{.1in}
\noindent\textbf{Unadjusted Estimator:}

\begin{align*}
\hat{ATT}_{unadj} &= \left\{E\left(Y(1) \mid A = 1\right) - E\left(Y(0) \mid A = 1\right)\right\} - \left\{E\left(Y(1) \mid A = 0\right) - E\left(Y(0) \mid A = 0\right)\right\} \\
&= \left\{\alpha_0 + \alpha_1 + \zeta_1 + \lambda_1 \tau_{11} + \gamma - (\alpha_0 + \alpha_1 + \zeta_0 + \lambda_0 \tau_{10}) \right\} - \\
& \ \ \ \  \left\{\alpha_0 + \zeta_1 + \lambda_1 \tau_{01} - (\alpha_0 + \zeta_0 + \lambda_0 \tau_{00}) \right\} \\
&= \gamma + \zeta_1 - \zeta_0 + \lambda_1 \tau_{11} - \lambda_0 \tau_{10} - (\zeta_1 - \zeta_0 + \lambda_1 \tau_{01} - \lambda_0 \tau_{00}) \\
&= \gamma + \lambda_1 \tau_{11} - \lambda_0 \tau_{10} - \lambda_1 \tau_{01} + \lambda_0 \tau_{00} \\
&= \gamma + \lambda_1(\tau_{11} - \tau_{01}) - \lambda_0 (\tau_{10} - \tau_{00}).
\end{align*}

Without significant restrictions on the $\lambda$ and $\tau$ values, this does not equal the ATT. \\

\vspace{.1in}
\noindent\textbf{Adjusted Estimator:}

\noindent Now, imagine we know which regression model to fit. In R, we can fit the model \verb|lm(y~a*t + x*t)|, which is correctly specified. The estimate of the treatment effect will be the coefficient on the interaction between \verb|a| (treatment indicator) and \verb|t| (time). However, when we fit the model, we will get:
\[
\hat{ATT}_{adj} = \gamma,
\]
which is biased for the true ATT.

\setcounter{table}{0}
\renewcommand{\thetable}{B\arabic{table}}

\section*{Appendix B - Calculation of ATT for Simulation Scenario 6} 

In the main paper, we state that the average treatment effect on the treated (ATT) in Scenario 6 is different than in the other scenarios. Here, we show our calculations for the ATT using our data-generating example. Below is the code used to generate data, using the \verb|dplyr| R package.

\begin{verbatim}
dat <- expand.grid(id = 1:n, tp = 1:max.time) %>% 
    arrange(id,tp) %>% group_by(id) %>%
    mutate(int=rnorm(1,0,sd=0.25), # random intercept
           p.trt=0.5, # probability of treatment
           trt=rbinom(1, 1, p.trt), # treatment
           x=rnorm(1, mean = 1.5 - 0.5*trt, sd = 1.5 - 0.5*trt),
           post=I(tp >= trt.time), # indicator of post-treatment period
           treated=I(post == 1 & trt == 1), # treated indicator
           x=ifelse(tp>=2, lag(x, 1) + (tp-1)/10 * 
                rnorm(1, mean = 1, sd = 0.1) - 
                I(trt == 1) * I(tp>6)*(tp)/20, x)
    ) %>% 
    ungroup()
  
dat <- dat %>% mutate(err=rnorm(n*max.time), 
                      y = 1 + x + trt + int + err + treated + 
		       ((tp - 2.5)^2)/10,
                      y.t =  1 + x * tp / 10 + trt +
                             int + err + treated + ((tp - 2.5)^2)/10) %>%
    group_by(id) %>%
    mutate(y.diff = y - lag(y), y.diff2 = y.t - lag(y.t)) %>% 
    ungroup()
\end{verbatim}

To begin, we \emph{only} need to look at the treated group since the ATT is defined on the treated population. The setup is relatively simple. We set $n = 1000$ to be the total number of units followed over 10 (\verb|max.time|) time points. Units were assigned to the treatment group with probability $0.5$. The treated units were given treatment beginning at $t = 6$; thus, we had five pre-treatment time points and five post-treatment time points. The covariate $X$ at baseline was drawn from a Normal distribution, $N(1, 1^2)$ from the treated population. During the pre-treatment period, the means of the covariate increased by about $\frac{1}{10}$ cumulatively from $t = 2, \dots, 10$. However, the mean of the covariate was affected by treatment too, so that for the treated group when $t \geq 6$, the mean went down by an average of $\frac{1}{20}$ per time point.

\begin{table}[!ht]
\begin{tabular}{lllllllllll}
Time  & $t = 1$ & $t = 2$ & $t = 3$ & $t = 4$ & $t = 5$  & $t = 6$ & $t = 7$ & $t = 8$ & $t = 9$ & $t = 10$ \\
\hline
Mean($X^0$) & 1.0 & 1.1 & 1.2 & 1.3 & 1.4 & 1.5 & 1.6 & 1.7 & 1.8 & 1.9   \\
Mean($X^1$) & 1.0 & 1.1 & 1.2 & 1.3 &  1.4 & 1.45 & 1.5 & 1.55 & 1.6 & 1.65 \\
\hline
\end{tabular}
\caption{Evolution of counterfactual means of covariate $X$ for the treated group.}
\end{table}

Note that for this simulation scenario, we have two different outcomes. In the first, denoted \verb|y|, the effect of $X$ on the outcome is the same at every time point. For the second outcome, denoted \verb|y.t|, the covariate has a time-varying effect on the outcome. The two outcome processes are detailed below:

\begin{align*}
y &= 1 + x + trt + int + err + treated + ((tp - 2.5)^2)/10 \\
y.t &= 1 + x * tp / 10 + trt + int + err + treated + ((tp - 2.5)^2)/10).
\end{align*}

So this difference is that in the second equation, $X$ interacts with time. Note that both $int$ and $err$ are mean zero normal random variables and $treated = 1$ whenever $tp >5$. (We are only considering the treated group. This would not be true for the comparison group.) Like we did for the mean of $X$, we can build a table for the means of $Y$ using the above equations.

For \verb|y|, we get the following results:

\begin{table}[H]
\scalebox{0.8}{
\begin{tabular}{lllllllllllll}
Time  & $t = 1$ & $t = 2$ & $t = 3$ & $t = 4$ & $t = 5$  & $t = 6$ & $t = 7$ & $t = 8$ & $t = 9$ & $t = 10$ & Avg. pre & Avg. post \\
\hline
Mean($Y^0$) & 3.225 & 3.125  & 3.225 & 3.525  & 4.025 & 4.725 & 5.625 & 6.725  & 8.025  & 9.525 & 3.425 & 6.925 \\
Mean($Y^1$) &  3.225 & 3.125  & 3.225 & 3.525  & 4.025 & 5.675  & 6.525 & 7.575  & 8.825  & 10.275 & 3.425 & 7.775 \\
\hline
\end{tabular}}
\caption{Evolution of counterfactual means of outcome $Y$ for the treated group.}
\end{table}

We'll calculate a few of these by hand to give an idea of what we're doing. Take the mean of $Y^0$ at $t = 7$:

\begin{align*}
y &= 1 + x + trt + int + err + treated + ((tp - 2.5)^2)/10 \\
&= 1 + x + 1 + 0 + 0 + 0 + (7 - 2.5)^2/10 \\ 
&= 1+1.6 + 1 + 0 + 0 + 0 + (7 - 2.5)^2/10 \\
&= 5.625.
\end{align*}
Here, we plugged in 1.6 for $x$ since it equals the untreated mean of the covariate (see Table B1). Both $int$ and $err$ are independent mean zero random variables so we plug in 0.

Following similar calculations, the mean of $Y^1$ at $t = 7$ is:

\begin{align*}
y &= 1 + x + trt + int + err + treated + ((tp - 2.5)^2)/10 \\
&= 1 + x + 1 + 0 + 0 + 1 + (7 - 2.5)^2/10 \\ 
&= 1 + 1.5 + 1 + 0 + 0 + 1 + (7 - 2.5)^2/10 \\
&= 6.525.
\end{align*}

The ATT here is $7.775 - 6.925 = 0.85$, which is calculated by taking the mean of the last 5 columns (the post-treatment time points) for each row and subtracting them.

And for \verb|y.t|, we get the following results:

\begin{table}[H]
\scalebox{0.8}{
\begin{tabular}{lllllllllllll}
Time  & $t = 1$ & $t = 2$ & $t = 3$ & $t = 4$ & $t = 5$  & $t = 6$ & $t = 7$ & $t = 8$ & $t = 9$ & $t = 10$ & Avg. pre & Avg. post\\
\hline
Mean($Y^0$) & 2.325 & 2.245 & 2.385 & 2.745 & 3.325 & 4.125 & 5.145 & 6.385 & 7.845 & 9.525 & 2.605 & 6.605 \\
Mean($Y^1$) & 2.325 & 2.245 & 2.385 & 2.745 & 3.325 & 5.095 & 6.075 & 7.265 & 8.665 & 10.275 & 2.605 & 7.475  \\
\hline
\end{tabular}}
\caption{Evolution of counterfactual means of outcome $Y$ for the treated group.}
\end{table}

The ATT here equals 0.87.

\end{document}